# Observing Monopoles in a Magnetic Analog of Ice

Michel J. P. Gingras
Department of Physics and Astronomy, University of Waterloo,
Waterloo, Ontario N2L 3G1, Canada,
&
Canadian Institute for Advanced Research/Quantum Materials Program,
Toronto, Ontario M5G 1Z8, Canada

*This is a nontechnical Perspective commentary on two recent neutron scattering experiments (Science **326**, 415 (2009) and Science **326**, 411 (2009)) reporting evidence for the validity of a "Coulomb phase" description of the low-temperature regime of frustrated spin ice magnetic materials as well as the existence of topological defect (monopole-like) excitations in these systems.*

A bar magnet has a north and south pole, and cutting it in half just creates two new poles, not two separated monopoles. However, a recent theoretical proposal suggested that defects in the spin alignment of certain oxide magnets can create separated effective magnetic monopoles [1]. These materials are called spin ices because the lowest-energy orientation of the magnetic spins closely mimics the most stable arrangement of protons in water ice (2). In two recent papers, Fennell *et al*. [3] and Morris *et al*. [4] report measurements from neutron-scattering experiments showing that the low-energy excitations in spin ices are reminiscent of Dirac's elementary magnetic monopoles [5] that have so far eluded the searches of high-energy physicists. These dissociated north and south poles diffuse away from each other [6] in these oxides and leave behind a "Dirac string" of reversed spins that can be seen as patterns in the intensity of scattered neutrons.

The materials studied, holmium titanate ($Ho_2Ti_2O_7$) [3] and dysprosium titanate ($Dy_2Ti_2O_7$) [4], are geometrically frustrated ferromagnets [7,8]; they are unusual paramagnets with strong spin-spin correlations [8] that become magnetized in a magnetic field. These oxides have a pyrochlore structure; the $Ho^{3+}$ and $Dy^{3+}$ ions within these crystals form a lattice of corner-sharing tetrahedra (see the figure, panel A). The magnetic moments of $Ho^{3+}$ and $Dy^{3+}$ act like two states, or Ising model spins, and are constrained by anisotropic forces to point "in" or "out" of the tetrahedra. The minimum energy condition, or rule, is that there must be two spins pointing "in" and two spins pointing "out" on each tetrahedron [7,8].

The "two-in, two-out" rule is analogous to the Bernal-Fowler ice rule, which states that for energetic reasons, two protons must be "close" and two protons must be "far" from any given oxygen in common water ice [2]. As explained by Pauling [2], the very large number of two-in, two-out configurations in a macroscopic ice sample leads to a measurable residual entropy in ice at low temperatures [9,10]. Correspondingly, $Ho_2Ti_2O_7$ and $Dy_2Ti_2O_7$ also exhibit a residual low-temperature "Pauling entropy" [11].

A tetrahedron fulfilling the two-in, two-out rule amounts to an effective "magnetic charge neutrality" where four (two positive, two negative) magnetic charges cancel out at the center of the tetrahedron [1]. A spin flipped by thermal fluctuations creates two defective adjoining tetrahedra—a monopole-antimonopole pair (see the figure, panel B). Once formed, these particles can diffuse away from each other by reversing spins along the path they trace as they separate, which reconstitutes tetrahedra that obey the ice rule along the path [6]. In his effort to alter the standard theory of electromagnetism as little as possible [12], Dirac [5] postulated that a monopole-antimonopole pair is connected by an unobservable "string" or tightly wound magnetic solenoids. In spin ices, the "Dirac strings" of reversed spins have direct consequences [3] and are observable [4]. Because all tetrahedra, except the two adjoining the north and south poles, fulfill the ice rule, the string is without tension and the energy to separate the poles by an infinite distance is finite [1] — the defects are dissociated, or deconfined.

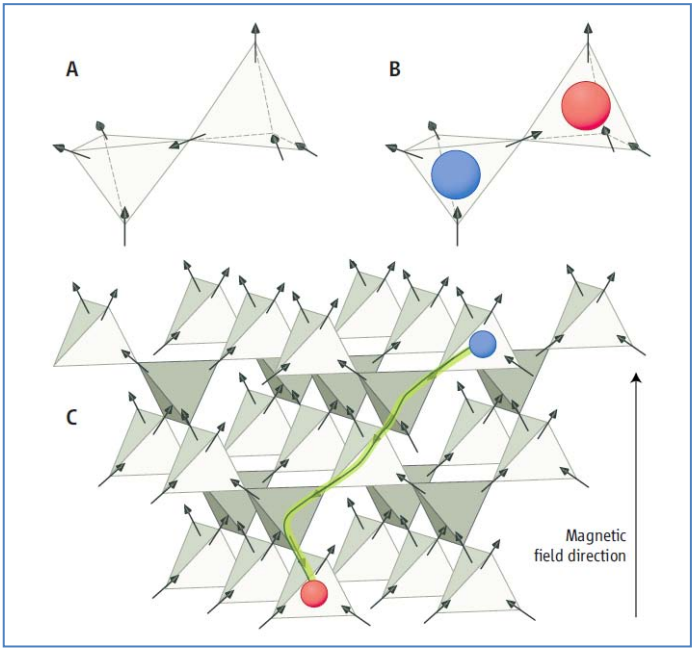

**Spin excitations creating magnetic monopoles.** (A) Spins on two adjacent tetrahedra for magnetic ions in the pyrochlore lattice satisfy a rule that requires two spins pointing in and two spins pointing out, similar to the arrangement of protons in water ice. (B) The reversal of a spin connecting two tetrahedra amounts to the creation of a "monopole" and an "antimonopole" that differ in magnetic "charge." At finite temperature, monopole-antimonopole pairs are created by thermal fluctuations. The monopoles can separate, leaving behind a "Dirac string" of reversed spins. Their signature was deduced in spin-polarized neutron scattering by Fennell *et al.* in zero applied magnetic fi eld, where the Dirac strings have no preferred orientation. (C) For a magnetic field applied along the crystallographic [100] direction shown, the ground state has the magnetic moments on each tetrahedron pointing with the field, which still maintains the two-in, two-out rule. Two separated tetrahedra, each with a flipped spin (three in, red; three out, blue) leads to a pair of monopoles connected by a string (green) of spins reversed against the field. Morris *et al.* observed a signature for this string in their neutron-scattering data by carefully tuning the applied magnetic field.

The monopoles in spin ice act like magnetic charges: They obey analogous electric field laws and exhibit an effective Coulomb's law for their interaction strength. At zero temperature, the spin-ice state can be viewed as a "vacuum" free of monopoles and is referred to as a "magnetic Coulomb phase" [3]. This analogy affords a mathematical framework for calculating the underlying spin correlations of the rare-earth magnetic moments. Thermal fluctuations that create dissociated monopoles are sources of the analogous electric field that modifies the spin correlations [3,4] and the dissociated monopoles can be used to describe the low-temperature thermodynamic properties of the material [1,4].

Neutron scattering can probe the "Coulomb phase" nature of the spin-ice state by measuring the spin-spin correlation function, $C(r)$, where $r$ is the distance between spins. In the absence of thermally induced monopoles, $C(r)$ does not decay exponentially with $r$, as would be the case for a conventional thermally disordered paramagnet. Rather, $C(r)$ is theoretically expected to display the same spatial anisotropy and $r^{-3}$ decay as a dipolar interaction. These correlations are manifest in the neutron scattering as "bow-tie" pinch-point singularities at particular neutron-scattering directions of wave vectors **Q**, which correspond to a "reciprocal space" of the real-space lattice in the crystals. The theoretical argument for the magnetic Coulomb phase [1] is highly compelling, but all previous neutron-scattering experiments, such as those on $Ho_2Ti_2O_7$ [13] and $Dy_2Ti_2O_7$ [14], failed to find an unmistakable signature of the pinch points. Unlike prior studies, Fennell *et al.* performed a polarized neutron-scattering experiment where the scattering signal is separated in two components. The pinch points are clearly revealed in the component where the neutron spin is flipped, confirming the theoretical prediction. The pinch points are obscured in the more intense "non-spin flip" signal, which helps to explain why previous studies were inconclusive.

In their analysis, Fennell *et al*. [3] introduced a parameter that cuts off the pinch-point singularities in reciprocal space, which they associate with the typical length of the Dirac strings. Whereas their experiment was performed in zero magnetic field and did not directly probe those strings, Morris *et al*. [4] applied a magnetic field **B** along the [*100*] crystal direction to induce a magnetically polarized state where the ice rule and the minimum magnetic field energy, or Zeeman energy, are satisfied simultaneously (see the figure, panel C). The magnetic field strength can be tuned near a transition where thermally excited monopole-antimonopole pairs start to proliferate. The resulting flipped spins of the Dirac string are then oriented against the magnetic field direction, with the strings causing cone-like features observable in the scattering intensity pattern. The conic features transform in inclined sheets of scattering when the field direction is tilted away from the [*100*] direction, in close concordance with the calculations of Morris *et al*. for this state. The specific heat in zero magnetic field can also be described well in terms of a dilute gas of thermally excited monopoles [4].

The demonstration that dissociated monopole-like excitations in spin ices can be observed and manipulated may help guide future studies of similar topological excitations in other exotic condensed matter systems. Of particular interest is the exploration of geometrically frustrated magnetic systems with large quantum mechanical zero-point fluctuations of the magnetic moments away from the classical Ising spin directions [15,16] enforced in the $Ho_2Ti_2O_7$ and $Dy_2Ti_2O_7$ spin ices considered in References [3,4]. Such quantum magnets could provide condensed matter physicists with systems that mimic the physics of quantum electrodynamics.